\documentclass[trackchanges,times]{aastex701}
\usepackage{times}

\begin{document}

\title{The ${}^{13}\mathrm{CO}(2{-}1)/^{12}\mathrm{CO}(2{-}1)$ Line Ratio from 100 Molecular Clouds in the Large Magellanic Cloud}

\author[orcid=0009-0000-4674-1195]{Dia Kalra}
\affiliation{Dept. of Physics, University of Alberta, 4-181 CCIS, Edmonton, AB T6G 2E1, Canada}
\email{dkalra1@ualberta.ca}

\author[orcid=0000-0002-5204-2259]{Erik Rosolowsky}
\affiliation{Dept. of Physics, University of Alberta, 4-181 CCIS, Edmonton, AB T6G 2E1, Canada}
\email{rosolowsky@ualberta.ca}

\author[0000-0002-2545-1700]{Adam Leroy}
\affiliation{Department of Astronomy, The Ohio State University, 140 West 18th Avenue, Columbus, OH 43210, USA}
\affiliation{Center for Cosmology and Astroparticle Physics (CCAPP), 191 West Woodruff Avenue, Columbus, OH 43210, USA}
\email{leroy.42@osu.edu}

\author[0000-0002-4663-6827]{Remy Indebetouw}
\affiliation{University of Virginia Astronomy Department, P.O. Box 400325, Charlottesville, VA, 22904, USA}
\affiliation{National Radio Astronomy Observatory, 520 Edgemont Rd, Charlottesville, VA 22903, USA}
\email{ri3e@virginia.edu}
\begin{abstract}
We analyze the line ratio of the \textsuperscript{13}CO (2-1) to \textsuperscript{12}CO (2-1) rotational transitions observed from new ALMA observations of 100 Giant Molecular Clouds (GMCs) that span the Large Magellanic Cloud. We measure a median line ratio of $^{13}\mathrm{CO}(2{-}1)/^{12}\mathrm{CO}(2{-}1) = 0.078$ with $68\%$ of the sample falling between 0.058 and 0.107. A regression analysis confirms a nearly linear relationship across two orders of magnitude in line luminosity. Moreover, we find that the inclusion of $(L_{\text{FIR}})$ from Young Stellar Objects as a predictor variable of the line ratio significantly improves the quality of the fit, with clouds hosting IR-bright YSOs having relatively brighter $^{13}$CO emission. This analysis indicates that active star forming molecular clouds have different internal conditions than more quiescent clouds.
\end{abstract}



\section{Introduction} 

Because of the invisibility of H$_2$ in cold molecular clouds, CO is the brightest tracer species used to map out the distribution of star-forming molecular gas \citep{araa_xco}. Several rotational transitions from different isotopologues of CO are detected in the Milky Way and extragalactically.  
%These different species probe the physical conditions of the molecular gas \citep[e.g.,][]{wilson1994}, which typically requires at least three different spectral line detections from at least two isotopologues. 
A single line ratio between a pair of lines does not yield a full solution for the conditions of the gas (density, temperature, relative abundance of the species), but single line ratios are proxies for changing conditions of the gas.  The ratio of individual spectral lines has been used to indicate the general excitation conditions of the molecular gas \citep[e.g., $^{12}\mathrm{CO}(2{-}1)/^{12}\mathrm{CO}(1{-}0)$;][]{leroy_2022} or a proxy for the typical density of the gas as inferred through its opacity \citep[e.g., $^{13}\mathrm{CO}(1{-}0)/^{12}\mathrm{CO}(1{-}0)$;][]{polk1988}.

Here, we measure the $R_{13}\equiv {}^{13}\mathrm{CO}(2{-}1)/^{12}\mathrm{CO}(2{-}1)$ line ratio for 100 molecular clouds in the Large Magellanic Cloud.  This line ratio, like its $(1{-}0)$ counterpart, traces the relative opacity of the two spectral lines.  The two lines should have comparable excitation temperatures \citep{mangum2015}, and would have $R_{13}$ set by the column density ratio \citep[$^{13}\mathrm{C}/^{12}\mathrm{C}\sim 1/50$ in previous LMC studies; e.g.,][]{johansson1994}.  Because of the opacity, $R_{13}$ is higher with the ratio varying based on the relative opacities of the two spectral lines. 

We report $R_{13}$ ratio primarily to measure its range of variation in an extragalactic environment and sensitivity to different molecular cloud properties. Notably, the Atacama Large Millimeter/submillimeter Array (ALMA) is capable of simultaneous observations of these two species in Band 6, so this line ratio will inform planned observations in that commonly used observing mode.

\section{Observations}
\begin{figure}
        \centering
        \plottwo{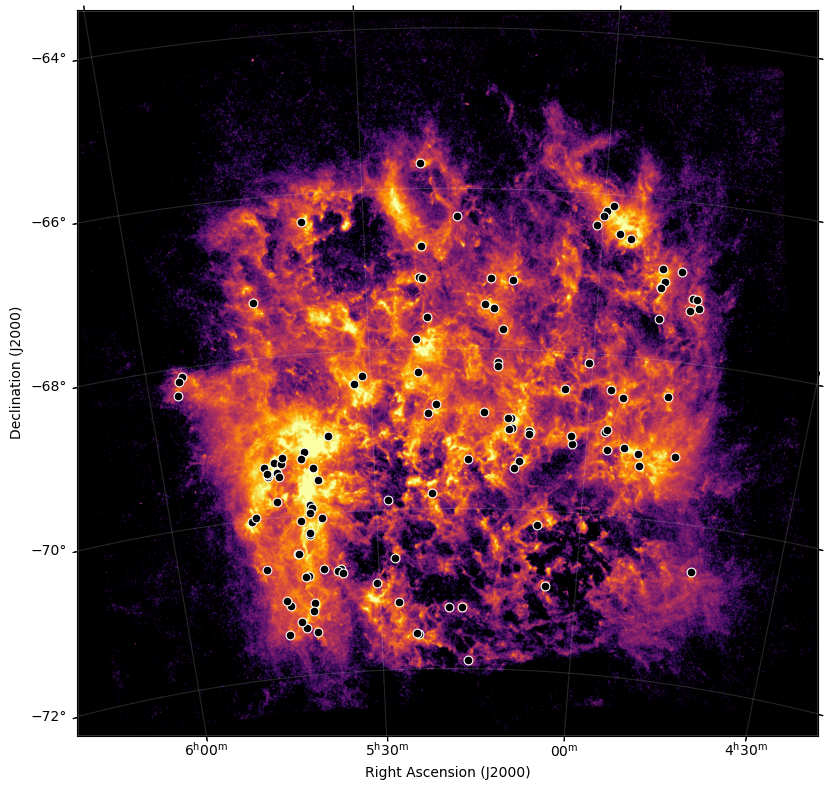}{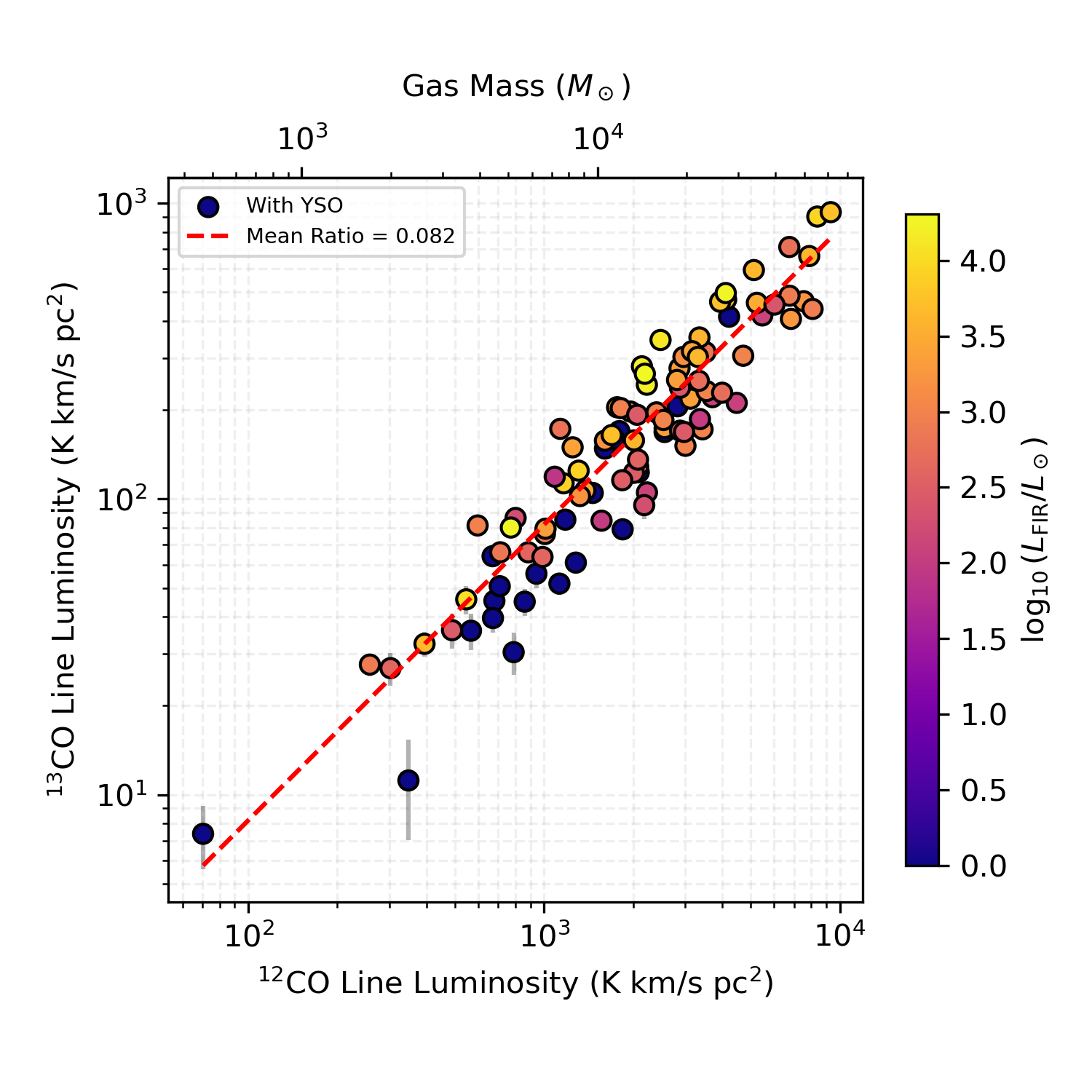}
        \caption{(left) Herschel SPIRE $500\,\mu$m dust continuum map of the LMC displayed on a logarithmic stretch. Circles indicate the positions of the ALMA fields.  (right) Correlation between  \textsuperscript{12}CO and  \textsuperscript{13}CO for 100 GMCs in the LMC. The points are color-coded by the far infrared luminosity $(\log L_{\text{FIR}})$ from YSOs inside the molecular clouds. The red dashed line represents the ensemble mean ratio ($R_{13}$ = 0.082).  The measurements for this figure are given in Table \ref{tab:data}}
        \label{fig:results}
        \digitalasset
\end{figure}

We analyze the pipeline-processed total power data from the ALMA projects 2023.1.00536.S, 2025.1.00589.S, which observed $^{12}$CO(2-1) and $^{13}$CO(2-1) from molecular clouds in the Large Magellanic Cloud (LMC). The targets were selected from detections the Magellanic Mopra Assessment (MAGMA) survey \citep{wong_magma}. The locations of the targets are indicated in Figure \ref{fig:results} (left).

\section{The $^{12}$CO to $^{13}$CO Line Ratio}

We measure $R_{13}$ by reprojecting the \textsuperscript{13}CO data cube onto the  \textsuperscript{12}CO position-position-velocity coordinate grid and integrating the emission using a common signal mask.  We generate a $4\sigma$ mask, expanded to connected $2\sigma$ using the methods described in \citet{phangs_pipeline}, where $\sigma$ is the noise level.  We then apply the same mask to the \textsuperscript{13}CO data. The spectral line luminosity is the sum of the data (in units of K) over this signal mask, multiplied by the projected volume of a pixel in the cube in units of $\mathrm{km~s^{-1}~pc}^2$.  The uncertainty in the luminosity derives from the mean error value within the mask and scaled by $\sqrt{N}$ where $N = n_\mathrm{pix} \Omega_\mathrm{pix}/\Omega_\mathrm{beam}$, where $n_\mathrm{pix}$ is the number of pixels in the 3D mask and $\Omega_\mathrm{pix}$ and $\Omega_\mathrm{beam}$ are the solid angles of a single pixel and the solid angle of the beam.  This implicitly assumes the spectral channels are statistically independent. These uncertainties are small compared to the measurements (typically $<3\%$), so our error budget is dominated by calibration uncertainties from ALMA \citep[$\sim 10\%$ at these frequencies][]{alma_guide}.  We also do not expressly correct for the 5\% difference between the angular resolution of the two species, but the clouds are resolved so this is likely negligible.

%it is unlikely we are missing significant $^{13}\mathrm{CO}$ emission outside the $^{12}\mathrm{CO}$ mask.

We also identify YSOs associated with the GMCs by finding all YSOs in the \citet{seale_ysos} catalog that fall within the spatial projection of the GMC mask. For each cloud, we measure  $L_\mathrm{FIR}$, the total far infrared luminosity for all sources associated with each GMC.

The GMCs in the sample have a median $\langle^{13}\mathrm{CO}/^{12}\mathrm{CO}\rangle =0.078$ with $68\%$ of the clouds falling within the range 0.058 - 0.107. The upper tail of this distribution remains well below the line ratio of $0.171 \pm 0.007$ derived by \citet{indebetouw2013}. This difference likely reflects both the broader cloud population sampled here and systematic differences because \citet{indebetouw2013} measure $R_{13}$ within \textsuperscript{13}CO-defined clump boundaries, whereas we apply a  \textsuperscript{12}CO-defined mask. 

Figure \ref{fig:results} (right) illustrates the relationship between cloud-integrated \textsuperscript{12}CO$(2{-}1)$ and \textsuperscript{13}CO$(2{-}1)$ intensities with color-coding showing the sum of the associated far infrared luminosity of YSOs. The data show a power-law relationship over two orders of magnitude. We fit a linear model to the data to measure the line ratio and assess the influence of the YSOs. We assess two models for predicting $\log_{10}{}^{13}\mathrm{CO}(2-1)$ intensity: one that depends only on \textsuperscript{12}CO and one that also includes $L_\mathrm{FIR}$. The single parameter linear fit implies a line ratio of ${}^{13}\mathrm{CO}(2{-}1)/{}^{12}\mathrm{CO}(2{-}1) = 0.082 \pm 0.002$. 

We compare the Bayesian Information Criterion (BIC) to compare the two models finding that including $L_\mathrm{FIR}$ reduces the BIC from $-94$ to $-122$.  Thus, including the effects of $L_\mathrm{FIR}$ produces a significantly better prediction of the $^{13}\mathrm{CO}$ luminosity.  The  best-fit relationship is: \begin{equation}
    \log L(^{13}\text{CO}) = (-1.31 \pm 0.08) + (0.92 \pm 0.03) \log L(^{12}\text{CO}) + (0.12 \pm 0.02) \log L_{\text{FIR}},
\end{equation}
where line luminosities are in units of K~km/s~pc$^{2}$ and $L_\mathrm{FIR}$ is in units of $L_\odot$. This model captures $91.4\%$ of the observed variance. The near unity slope shows that clouds have an approximately constant line ratio but the coefficient for luminosity indicates that, for a fixed \textsuperscript{12}CO intensity, clouds with higher $L_\mathrm{FIR}$ have higher \textsuperscript{13}CO emission. This cause of this increased line ratio could reflect that clouds with higher densities have more active star formation or it could indicate that IR heating is increasing the excitation of both species.  
%The resulting trend appears consistent with previous studies at small scales in the Milky Way \citep{sun10}, though the results of large-scale extragalactic studies imply the opposite relationship where regions with substantially higher star formation intensities show a lower $R_{13}$ \citep{davis2014}.

% The physical contrast between the two cloud populations is umambiguous. Quiescent clouds are confined to the low-intensity regime and sit below the mean-ratio line, consistent with sub-thermal \textsuperscript{13}CO excitation at low column densities and the absence of internal radiative heating. Conversely, the most infrared-luminous clouds anchor the high-intensity end and exceed gas-only prediction by up to a factor of 2. This discrepancy is fully recovered by the two-parameter model and is attributable to protostellar feedback thermally amplifying the dense gas emissivity. Taken together, these results establish that \textsuperscript{12}CO traces the bulk molecular gas reservoir while $L_{\text{FIR}}$  captures the luminosity-dependent enhancement driven by embedded star formation.

%% Please use the acknowledgment and contribution environments. This will 
%% be anonomyized when the "anonymous" style option is used. 
\begin{acknowledgments}
This paper makes use of the following ALMA data: ADS/JAO.ALMA\#2023.1.00536.S and ADS/JAO.ALMA\#2025.1.00589.S. ALMA is a partnership of ESO, NSF (USA) and NINS (Japan), together with NRC (Canada), NSTC and ASIAA (Taiwan), and KASI (Republic of Korea), in cooperation with the Republic of Chile. DK acknowledges the use Google Gemini \citep{gemini} to refine the phrasing of individual sentences and assist with debugging the analysis code.
\end{acknowledgments}

\begin{contribution}
%%This section gives authors the space to recognize author contributions. The text inside this environment is NOT counted towards the total word quanta. At a minimum, manuscripts are expected to include this text:

DK and ER analyzed the data and carried out the primary analysis.  All authors contributed to the writing and scientific development of this contribution.

%% But authors are expected to provide more specific details, e.g. 
%%
%%SC was responsible for writing and submitting the manuscript.
%%WWM came up with the initial research concept and edited the manuscript.
%%OTS obtained the funding and edited the manuscript.
%%EBF provided the formal analysis and validation. He also edited the manuscript.
%%GEH Supervised the undergraduates, wrote the software and administers the project github and Zenodo repositories.
%%
%% Authors can use the Contributor Role Taxonomy (CRediT) at
%% for ideas on how write a good statement tailored to their needs.

\end{contribution}

\facilities{ALMA}

\software{astropy \citep{2013A&A...558A..33A,2018AJ....156..123A,2022ApJ...935..167A}, spectral-cube \citep{spectral_cube}, ALMA pipeline \citep{alma_pipeline}}


\bibliography{sample701}{}
\bibliographystyle{aasjournalv7}


\appendix
\startlongtable
\begin{deluxetable}{lrrrrrrrrrrr}
\tablecolumns{11}
\tabletypesize{\footnotesize}
\tablecaption{\label{tab:data} Spectral line luminosities of molecular clouds in the Large Magellanic Cloud.  The Cloud ID is the object identifier used in the ALMA Archive. Luminosity measurements and the line ratio are given as $L$ and $R_{13}$ respectively with $1\sigma$ uncertainties given as $\delta L$ and $\delta R$.  The YSO flag indicates an association with at least one YSO as tabulated in \citet{seale_ysos}.  The Far IR luminosity of all associated YSOs is given as $L_\mathrm{FIR}$.}
\tablehead{
\colhead{Cloud ID} & \colhead{RA} & \colhead{Dec} & 
\colhead{$L_\mathrm{^{12}CO}$} & \colhead{$\delta L_\mathrm{^{12}CO}$} & 
\colhead{$L_\mathrm{^{13}CO}$} & \colhead{$\delta L_\mathrm{^{13}CO}$} & 
\colhead{$R_{13}$} & \colhead{$\delta R_{13}$} &
\colhead{YSO Flag} & \colhead{$L_\mathrm{FIR}$} \\
& \colhead{(deg)} & \colhead{(deg)} &
\colhead{($\mathrm{K\,km/s\,pc^2}$)} & 
\colhead{($\mathrm{K\,km/s\,pc^2}$)} & 
\colhead{($\mathrm{K\,km/s\,pc^2}$)} &
\colhead{($\mathrm{K\,km/s\,pc^2}$)} &
& & & \colhead{($L_\odot$)}
}
\startdata
LMCGMC\_021 & 86.70967 & -70.6318 &  89.3 & 2.4 & -0.4 & 2.4 &-0.004 & 0.027 & FALSE&     0 \\ 
LMCGMC\_105 & 86.41104 & -69.3712 & 347.9 & 4.1 & 11.2 & 4.2 & 0.032 & 0.012 & TRUE &     0 \\ 
LMCGMC\_136 & 85.16446 & -70.7618 &1461.0 & 7.6 &104.8 & 7.6 & 0.072 & 0.005 & TRUE &     0 \\ 
LMCGMC\_466 & 72.99212 & -67.0973 &1797.0 & 6.7 &169.6 & 6.8 & 0.094 & 0.004 & TRUE &     0 \\ 
LMCGMC\_474 & 74.91362 & -66.2061 &   0.0 & 0.1 &  0.0 & 0.1 &-0.058 & 1.564 & FALSE&     0 \\ 
LMCGMC\_072 & 77.75129 & -67.1381 &  70.3 & 1.7 &  7.4 & 1.8 & 0.105 & 0.025 & TRUE &     0 \\ 
LMCGMC\_115 & 81.78333 & -71.1537 & 679.3 & 4.2 & 45.3 & 4.3 & 0.067 & 0.006 & TRUE &     0 \\ 
LMCGMC\_161 & 85.95558 & -69.4501 &4222.5 & 9.9 &413.4 &10.0 & 0.098 & 0.002 & TRUE &     0 \\ 
LMCGMC\_255 & 84.95742 & -69.2164 &2825.7 & 9.8 &205.8 & 9.8 & 0.073 & 0.003 & TRUE &     0 \\ 
LMCGMC\_364 & 74.37892 & -68.9627 &1006.5 & 4.9 & 76.1 & 5.0 & 0.076 & 0.005 & TRUE &   900 \\ 
LMCGMC\_395 & 75.52487 & -69.1431 & 302.5 & 3.3 & 26.8 & 3.4 & 0.089 & 0.011 & TRUE &   410 \\ 
LMCGMC\_434 & 75.18438 & -66.3956 &1308.9 & 5.3 &124.7 & 5.4 & 0.095 & 0.004 & TRUE &  8610 \\ 
LMCGMC\_447 & 77.01358 & -69.0194 &3006.7 & 9.3 &151.3 & 8.9 & 0.050 & 0.003 & TRUE &  1050 \\ 
LMCGMC\_153 & 84.88338 & -69.8880 &2555.8 &10.8 &168.4 &10.2 & 0.066 & 0.004 & TRUE &     0 \\ 
LMCGMC\_312 & 71.71142 & -67.3062 & 566.1 & 5.3 & 35.9 & 5.1 & 0.063 & 0.009 & TRUE &     0 \\ 
LMCGMC\_313 & 71.94263 & -67.1954 &2090.3 & 8.6 &123.1 & 8.3 & 0.059 & 0.004 & TRUE &     0 \\ 
LMCGMC\_413 & 72.90363 & -67.0204 & 789.1 & 5.1 & 30.3 & 4.9 & 0.038 & 0.006 & TRUE &     0 \\ 
LMCGMC\_416 & 77.02367 & -69.0174 &3435.0 &11.7 &172.0 &10.7 & 0.050 & 0.003 & TRUE &  1050 \\ 
LMCGMC\_178 & 80.46313 & -69.8096 &1134.6 & 5.5 &172.6 & 5.8 & 0.152 & 0.005 & TRUE &   600 \\ 
LMCGMC\_198 & 70.71779 & -70.5685 & 545.6 & 4.9 & 45.8 & 5.1 & 0.084 & 0.009 & TRUE & 11120 \\ 
LMCGMC\_272 & 73.68396 & -69.1364 & 596.2 & 5.6 & 81.4 & 5.9 & 0.136 & 0.010 & TRUE &   980 \\ 
LMCGMC\_467 & 73.01900 & -66.8615 & 488.5 & 4.7 & 36.0 & 4.9 & 0.074 & 0.010 & TRUE &   280 \\ 
LMCGMC\_037 & 86.29825 & -69.4890 &2227.9 & 8.5 &105.2 & 8.5 & 0.047 & 0.004 & TRUE &   130 \\ 
LMCGMC\_417 & 78.05137 & -67.7514 & 942.1 & 5.8 & 55.9 & 5.9 & 0.059 & 0.006 & TRUE &     0 \\ 
LMCGMC\_439 & 75.01242 & -66.2689 &1164.5 & 5.5 &113.0 & 5.6 & 0.097 & 0.005 & TRUE &  8070 \\ 
LMCGMC\_542 & 80.70850 & -66.7177 & 669.7 & 3.9 & 64.1 & 4.0 & 0.096 & 0.006 & TRUE &     0 \\ 
LMCGMC\_390 & 80.77517 & -67.1039 &3515.7 & 8.1 &313.7 & 8.2 & 0.089 & 0.002 & TRUE &   680 \\ 
LMCGMC\_485 & 80.54496 & -68.8064 &2871.0 & 7.8 &276.7 & 7.9 & 0.096 & 0.003 & TRUE &  1990 \\ 
LMCGMC\_132 & 85.27188 & -70.7773 &3709.2 & 9.7 &220.9 & 9.9 & 0.060 & 0.003 & TRUE &   120 \\ 
LMCGMC\_320 & 83.05454 & -68.4142 &2557.1 & 8.3 &173.7 & 8.3 & 0.068 & 0.003 & TRUE &  2100 \\ 
LMCGMC\_225 & 77.67029 & -68.8526 &2013.8 & 5.9 &157.8 & 5.8 & 0.078 & 0.003 & TRUE &  3970 \\ 
LMCGMC\_298 & 86.04396 & -69.3157 &3128.4 & 8.3 &218.4 & 8.0 & 0.070 & 0.003 & TRUE &  2270 \\ 
LMCGMC\_350 & 80.86929 & -68.2928 &1564.4 & 6.7 & 84.4 & 6.7 & 0.054 & 0.004 & TRUE &    96 \\ 
LMCGMC\_368 & 75.86104 & -68.4657 &1766.0 & 5.4 &204.3 & 5.5 & 0.116 & 0.003 & TRUE &  4730 \\ 
LMCGMC\_229 & 79.16054 & -69.3861 &2881.5 & 7.4 &237.7 & 7.4 & 0.082 & 0.003 & TRUE &   240 \\ 
LMCGMC\_331 & 75.14021 & -68.1190 & 883.8 & 3.8 & 66.0 & 3.9 & 0.075 & 0.004 & TRUE &   390 \\ 
LMCGMC\_484 & 85.07825 & -69.2961 &2400.5 & 7.2 &196.3 & 7.2 & 0.082 & 0.003 & TRUE &   940 \\ 
LMCGMC\_547 & 78.64800 & -67.4385 &1606.8 & 5.0 &148.1 & 5.2 & 0.092 & 0.003 & TRUE &     0 \\ 
LMCGMC\_074 & 82.59088 & -70.9072 &1248.9 & 4.0 &149.5 & 4.1 & 0.120 & 0.003 & TRUE &  2430 \\ 
LMCGMC\_458 & 88.80575 & -68.1055 &1086.6 & 4.4 &118.9 & 4.5 & 0.109 & 0.004 & TRUE &    80 \\ 
LMCGMC\_523 & 78.34321 & -67.4829 &1726.3 & 5.6 &159.6 & 5.7 & 0.092 & 0.003 & TRUE &     0 \\ 
LMCGMC\_528 & 79.55700 & -66.3475 &1960.8 & 5.4 &197.5 & 5.5 & 0.101 & 0.003 & TRUE &  1800 \\ 
LMCGMC\_152 & 84.82688 & -69.9213 &7549.8 &14.7 &466.2 &14.4 & 0.062 & 0.002 & TRUE &  1720 \\ 
LMCGMC\_271 & 85.79733 & -69.3393 &1606.2 & 6.0 &157.4 & 6.1 & 0.098 & 0.004 & TRUE &  1610 \\ 
LMCGMC\_266 & 85.00208 & -70.2515 &4713.7 &12.4 &305.0 &12.4 & 0.065 & 0.003 & TRUE &  1000 \\ 
LMCGMC\_321 & 82.78350 & -68.3125 &1180.1 & 6.2 & 85.1 & 6.3 & 0.072 & 0.005 & TRUE &     0 \\ 
LMCGMC\_008 & 84.96746 & -70.2335 &3541.3 &10.3 &231.8 &10.0 & 0.065 & 0.003 & TRUE &  1000 \\ 
LMCGMC\_402 & 84.05225 & -69.0349 &8089.6 &16.2 &439.2 &15.7 & 0.054 & 0.002 & TRUE &   850 \\ 
LMCGMC\_019 & 85.00483 & -71.4711 &2181.3 & 9.8 & 95.3 &10.0 & 0.044 & 0.005 & TRUE &   192 \\ 
LMCGMC\_292 & 84.67142 & -69.4182 &2142.1 & 7.3 &280.9 & 7.6 & 0.131 & 0.004 & TRUE & 16440 \\ 
LMCGMC\_098 & 79.08054 & -71.8921 &2815.8 & 7.8 &252.6 & 8.0 & 0.090 & 0.003 & TRUE &  2200 \\ 
LMCGMC\_121 & 81.02421 & -71.5695 &1378.4 & 6.7 &106.8 & 6.9 & 0.077 & 0.005 & TRUE &  2810 \\ 
LMCGMC\_192 & 73.10662 & -69.3496 &1817.2 & 6.0 &202.8 & 5.9 & 0.112 & 0.003 & TRUE &   940 \\ 
LMCGMC\_352 & 85.01058 & -71.1055 &5469.3 &11.0 &417.4 &10.8 & 0.076 & 0.002 & TRUE &   130 \\ 
LMCGMC\_189 & 81.86479 & -70.6075 &2960.6 & 9.5 &302.2 & 9.4 & 0.102 & 0.003 & TRUE &  1390 \\ 
LMCGMC\_475 & 80.55508 & -67.6061 &1693.5 & 5.3 &164.5 & 5.4 & 0.097 & 0.003 & TRUE &  5300 \\ 
LMCGMC\_425 & 72.97454 & -67.4928 &1322.7 & 6.1 &102.3 & 6.1 & 0.077 & 0.005 & TRUE &  1680 \\ 
LMCGMC\_517 & 78.46392 & -67.1202 &1011.4 & 4.7 & 79.3 & 4.6 & 0.078 & 0.005 & TRUE &  2200 \\ 
LMCGMC\_252 & 73.18412 & -69.1931 &2226.3 & 7.4 &243.4 & 7.4 & 0.109 & 0.003 & TRUE & 16360 \\ 
LMCGMC\_297 & 85.90804 & -69.4960 & 987.9 & 5.1 & 63.7 & 5.1 & 0.064 & 0.005 & TRUE &   410 \\ 
LMCGMC\_022 & 84.57533 & -70.6890 &2476.4 & 6.9 &344.9 & 7.0 & 0.139 & 0.003 & TRUE & 13150 \\ 
LMCGMC\_188 & 77.65000 & -68.9872 & 801.3 & 3.8 & 86.3 & 3.9 & 0.108 & 0.005 & TRUE &   190 \\ 
LMCGMC\_014 & 85.41592 & -71.4026 & 671.8 & 4.2 & 39.6 & 4.2 & 0.059 & 0.006 & TRUE &     0 \\ 
LMCGMC\_103 & 85.27567 & -70.0682 &3352.6 & 8.1 &351.7 & 8.2 & 0.105 & 0.002 & TRUE &  3850 \\ 
LMCGMC\_450 & 74.11779 & -66.5361 &2191.4 & 6.9 &265.1 & 6.8 & 0.121 & 0.003 & TRUE & 20200 \\ 
LMCGMC\_489 & 78.18988 & -68.1595 &2080.1 & 8.7 &129.0 & 8.5 & 0.062 & 0.004 & TRUE &   956 \\ 
LMCGMC\_141 & 86.29662 & -69.4642 &1126.8 & 4.5 & 51.7 & 4.4 & 0.046 & 0.004 & TRUE &     0 \\ 
LMCGMC\_149 & 77.34579 & -69.3919 &3163.1 & 7.8 &316.1 & 7.7 & 0.100 & 0.002 & TRUE &  2800 \\ 
LMCGMC\_216 & 77.75825 & -68.9980 & 708.2 & 4.2 & 50.7 & 4.1 & 0.072 & 0.006 & TRUE &     0 \\ 
LMCGMC\_108 & 77.53383 & -69.4810 & 394.3 & 3.1 & 32.4 & 3.1 & 0.082 & 0.008 & TRUE &  4770 \\ 
LMCGMC\_232 & 77.04967 & -69.0450 &6834.9 &13.8 &406.4 &13.5 & 0.059 & 0.002 & TRUE &  1790 \\ 
LMCGMC\_133 & 86.08683 & -69.8077 & 711.1 & 3.4 & 66.0 & 3.4 & 0.093 & 0.005 & TRUE &   720 \\ 
LMCGMC\_353 & 84.54612 & -69.5793 &7881.6 &12.4 &662.8 &12.4 & 0.084 & 0.002 & TRUE &  4100 \\ 
LMCGMC\_055 & 87.07925 & -70.0204 & 772.2 & 4.1 & 80.1 & 4.1 & 0.104 & 0.005 & TRUE & 18000 \\ 
LMCGMC\_090 & 79.83283 & -71.2326 &2064.9 & 7.5 &192.6 & 7.6 & 0.093 & 0.004 & TRUE &   288 \\ 
LMCGMC\_148 & 86.33908 & -69.4474 &1280.7 & 4.9 & 60.9 & 4.9 & 0.048 & 0.004 & TRUE &     0 \\ 
LMCGMC\_343 & 73.89375 & -68.5184 &2012.7 & 7.6 &122.6 & 7.6 & 0.061 & 0.004 & TRUE &   300 \\ 
LMCGMC\_207 & 83.92488 & -70.7082 &2883.3 & 9.7 &170.2 & 9.5 & 0.059 & 0.003 & TRUE &   333 \\ 
LMCGMC\_539 & 84.41812 & -66.3432 & 257.6 & 2.5 & 27.5 & 2.4 & 0.107 & 0.009 & TRUE &   810 \\ 
LMCGMC\_283 & 84.48629 & -70.0522 & 860.0 & 4.7 & 44.9 & 4.6 & 0.052 & 0.005 & TRUE &     0 \\ 
LMCGMC\_544 & 80.69129 & -67.1200 &5243.0 &10.6 &460.5 &10.6 & 0.088 & 0.002 & TRUE &  3200 \\ 
LMCGMC\_199 & 85.50458 & -70.4746 &1837.8 & 6.7 &115.7 & 6.4 & 0.063 & 0.003 & TRUE &   320 \\ 
LMCGMC\_279 & 82.06142 & -69.8836 &5121.6 & 7.9 &594.3 & 7.8 & 0.116 & 0.002 & TRUE &  4200 \\ 
LMCGMC\_020 & 85.95696 & -71.1093 &1844.3 & 6.0 & 78.9 & 5.8 & 0.043 & 0.003 & TRUE &     0 \\ 
LMCGMC\_282 & 71.79333 & -67.2009 &4480.6 &12.3 &211.3 &12.1 & 0.047 & 0.003 & TRUE &   117 \\ 
LMCGMC\_387 & 75.56812 & -69.0456 &2531.9 & 6.6 &184.9 & 6.6 & 0.073 & 0.003 & TRUE &  1020 \\ 
LMCGMC\_214 & 71.86583 & -69.1743 &3311.4 & 8.1 &302.2 & 8.1 & 0.091 & 0.002 & TRUE &  4430 \\ 
LMCGMC\_543 & 80.68271 & -65.6824 &6748.9 & 9.7 &711.3 & 9.7 & 0.105 & 0.001 & TRUE &   580 \\ 
LMCGMC\_422 & 74.45325 & -66.4824 &8392.9 &11.2 &901.4 &11.1 & 0.107 & 0.001 & TRUE &  8630 \\ 
LMCGMC\_179 & 74.27621 & -69.1909 &4127.2 & 9.8 &472.0 & 9.8 & 0.114 & 0.002 & TRUE &  3730 \\ 
LMCGMC\_125 & 85.45308 & -70.4748 &6757.9 &13.6 &486.7 &13.3 & 0.072 & 0.002 & TRUE &   830 \\ 
LMCGMC\_379 & 80.28696 & -68.6903 &3337.4 & 8.2 &250.9 & 8.2 & 0.075 & 0.002 & TRUE &   510 \\ 
LMCGMC\_340 & 74.31112 & -68.4391 &3937.7 & 9.7 &464.4 & 9.7 & 0.118 & 0.002 & TRUE &  5570 \\ 
LMCGMC\_264 & 77.79063 & -68.8553 &6012.5 & 8.9 &454.2 & 8.8 & 0.076 & 0.001 & TRUE &   238 \\ 
LMCGMC\_006 & 85.08788 & -71.2104 &9317.3 &11.8 &932.8 &11.6 & 0.100 & 0.001 & TRUE &  5610 \\ 
LMCGMC\_184 & 84.06121 & -70.7330 &3365.5 & 9.5 &186.2 & 9.5 & 0.055 & 0.003 & TRUE &   100 \\ 
LMCGMC\_482 & 78.16400 & -68.2158 &2078.1 & 7.7 &135.8 & 7.7 & 0.065 & 0.004 & TRUE &   376 \\ 
LMCGMC\_381 & 83.87400 & -70.7601 &2974.1 &10.2 &168.5 &10.2 & 0.057 & 0.003 & TRUE &   289 \\ 
LMCGMC\_043 & 79.35488 & -71.2378 &4115.0 & 8.5 &496.9 & 8.6 & 0.121 & 0.002 & TRUE & 17370 \\ 
LMCGMC\_018 & 86.11400 & -71.4720 &4000.9 &12.4 &228.7 &12.1 & 0.057 & 0.003 & TRUE &   525 \\ 
\enddata
\end{deluxetable}

\end{document}